\documentclass[12pt,cite,epsf,epsfig]{article}
\usepackage{epsfig}
\title{Model Independent Constraints on Non-electronic Flavors in the Solar Boron Neutrino Flux}
\author{S. Dev\thanks{E--mail : dev5703@yahoo.com}, Sanjeev Kumar\thanks{E--mail : sanjeev3kumar@yahoo.co.in } and Surender Verma\\
	{\em Department of Physics, Himachal Pradesh University,} \\
	 {\em Shimla, India-171005.}}

\begin{document}
\maketitle

\begin{abstract}
We perform the most general model independent analysis of the latest 391-Day
Salt Phase SNO Data Set incorporating the Super-Kamiokande ES flux
measurement and obtain bounds on the antineutrino and sterile neutrino flux
in the solar $^{8}$B neutrino flux reaching the detectors on the earth. The
muon/ tauon antineutrino flux is found to be disallowed at 1.4$\sigma $ C.L.
The sterile flux is found to be non-zero at about 1.26 standard deviations.
\end{abstract}

The electron neutrinos coming from the Sun were first detected through their
Charge Current (CC) interactions with Chlorine \cite{1}. The CC reactions
detect electron neutrinos only since they are insensitive to the neutrinos
of other flavors. The $\nu _{e}$ fluxes measured in different solar neutrino
experiments \cite{1,2} were found to be smaller than the corresponding
Standard Solar Model (SSM) \cite{3} estimates giving rise to the solar
neutrino problem (SNP). The measurement of solar $^{8}$B neutrino flux
through the elastic scattering (ES) reaction at Super-Kamiokande \cite{4}
and through the neutral current (NC) reaction at SNO \cite{5} were two major
steps towards the resolution of the SNP. The ES flux measurement of SNO
agrees with the ES flux measurement at Super-Kamiokande even though the
latter have lesser precision. The ES flux at Super-Kamiokande was found to
be larger than the SNO CC flux. Interpreting this excess as the contribution
of non- electronic neutrino flux resulting from the flavor conversion of
electron neutrinos, the hypothesis of the oscillation of solar $^{8}$B
neutrinos was, indirectly, established at 3.2$\sigma $ C.L.\cite{4}. A more
direct evidence for the flavor conversion of solar $^{8}$B neutrinos was
provided by the NC flux measurement of the boron neutrino flux at SNO \cite
{5} which was consistent with the SSM prediction for the solar $^{8}$B
neutrino flux. This was followed by the independent observation of the
oscillations of the terrestrial antineutrinos at KamLAND \cite{6} with the
LMA parameters. After these pioneering experiments, neutrino masses and,
consequent, oscillations have been established as a physical reality. After
the observation of neutrino masses through oscillations, neutrino magnetic
moments are an inevitable consequence in the standard model (SM) and beyond.

The KamLAND results, directly, ruled out many other possible solutions of
the SNP including the resonant spin flavor precession (RSFP) solution
resulting from the interaction of a non-vanishing neutrino magnetic moment
with the solar magnetic fields. However, the presently available solar
neutrino data do not rule out the subdominant contributions of the spin
flavor precession (SFP) accompanying the dominant LMA flavor conversion. The
SFP is very small for the neutrino parameters in the LMA region but the SFP
conversion could be, significantly, enhanced under suitable conditions \cite
{7}. A particularly interesting scenario \cite{8} has, recently, been
discussed in the literature with non-vanishing transition magnetic moment in
the $\mu $-$\tau $ sector. In this scenario, the final solar neutrino flux
can be a mixture of $\nu _{e}$, $\nu _{\mu }$, $\nu _{\tau }$, $\overline{%
\nu }_{\mu }$, $\overline{\nu }_{\tau }$ and there will be a direct
correlation between the relative fluxes of different neutrino families and
the solar magnetic fields. Therefore, it is of considerable importance to
constrain the probable non-electronic solar antineutrino flux in a model
independent manner keeping, also, in mind the, rather, stringent constraints
on the solar electron antineutrino flux \cite{9} since the possible presence
of these effects, even at the subdominant level, may affect our present
knowledge of the neutrino masses and mixings. Consequently, it is of utmost
importance to constrain the possible presence of these subdominant effects
from the analysis of solar and atmospheric neutrino data. In the present
work, we intend to constrain the possible presence of subdominant SFP
transitions and transitions into sterile neutrinos in addition to the
dominant LMA transitions in a model independent manner so that the
constraints derived in the present work not only apply to the usual SFP
transitions \cite{10} but, also, to the more exotic SFP scenarios studied in
the recent past \cite{7,8}.

The possibility of transitions into active antineutrinos and sterile
neutrinos accompanying the dominant LMA transitions has been examined by
several authors \cite{11,12} in the recent past through a model independent
analysis of the Super-Kamiokande and SNO data. In the present work, we
reexamine this possibility in the light of the latest 391-Day Salt Phase SNO
Data Set and the Super-Kamiokande measurement of the $^{8}$B flux through
the ES reaction.

If the solar $^{8}$B neutrinos undergo flavor conversion into active
neutrinos only and there are no transitions into antineutrinos, the CC, NC
and ES fluxes are no longer independent. The ES flux can be obtained from
the CC and NC fluxes by using the relation \cite{13}
\begin{equation}
\phi _{ES}=r\phi _{NC}+(1-r)\phi _{CC},
\end{equation}
where `r' is the ratio of the ES cross-section of the $\nu _{\mu }/\nu
_{\tau }$ component to that of $\nu _{e}$. This relation should be satisfied
by the global rates as well as in the individual energy bins. Moreover, Eqn.
(1) gives rise to correlations between the CC, NC and ES fluxes which
should, also, be visible in the SNO data. However, these correlations will
be modified in case there are transitions into antineutrinos at the
subdominant level. Therefore, the correlations between different SNO fluxes
should serve as the test of the LMA MSW solution and may, also, indicate the
presence of subdominant transitions.

In this work, we make use of the 391-Day Salt Phase SNO Data Set with
enhanced sensitivity. The shape constrained SNO CC, NC and ES fluxes are
given by \cite{14}
\begin{equation}
\phi _{CC}^{SNO}=\left( 1.72\pm 0.12\right) \times 10^{6} cm^{-2}s^{-1},
\end{equation}
\begin{equation}
\phi _{NC}^{SNO}=\left( 4.81\pm 0.34\right) \times 10^{6} cm^{-2}s^{-1},
\end{equation}
\begin{equation}
\phi _{ES}^{SNO}=\left( 2.34\pm 0.27\right) \times 10^{6} cm^{-2}s^{-1}.
\end{equation}
The correlation coefficients are given by \cite{14}
\begin{equation}
\rho \left( CC,NC\right) =-0.400,
\end{equation}
\begin{equation}
\rho \left( CC,ES\right) =-0.168,
\end{equation}
\begin{equation}
\rho \left( ES,NC\right) =-0.073,
\end{equation}
and will be used in the error propagation calculations.

The starting point of our analysis are the relations between the CC, NC and
ES fluxes which are measured experimentally and the neutrino/ antineutrino
fluxes of various species which may be present in the solar $^{8}$B neutrino
flux. We do not parametrize the contribution of the various fluxes in terms
of some trigonometric ratios representing the fractions of different
neutrino fluxes as is done in Ref. [11] and Ref. [12]. Our approach,
although equivalent, is much simpler and more transparent.

The CC, NC and ES fluxes are related to the various neutrino fluxes in a
most general way through the relations
\begin{equation}
\phi _{CC}=\phi _{\nu _{e}},
\end{equation}
\begin{equation}
\phi _{NC}=\phi _{\nu _{e}}+\phi _{\nu _{x}}+\overline{r}_{d}\phi _{%
\overline{\nu }_{e}}+\overline{r}_{d}\phi _{\overline{\nu }_{x}},
\end{equation}
\begin{equation}
\phi _{ES}=\phi _{\nu _{e}}+r\phi _{\nu _{x}}+\overline{r}_{_{e}}\phi _{%
\overline{\nu }_{e}}+\overline{r}_{x}\phi _{\overline{\nu }_{x}}.
\end{equation}
Here, $\phi _{\nu _{e}}$ is the flux of $\nu _{e}$, $\phi _{\nu _{x}}$ is
the flux of $\nu _{\mu }/\nu _{\tau }$, $\phi _{\overline{\nu }_{e}}$ is the
flux of $\overline{\nu }_{e}$ and $\phi _{\overline{\nu }_{x}}$ is the flux
of $\overline{\nu }_{\mu }/\overline{\nu }_{\tau }$. We have assumed that
all these neutrinos come with standard boron neutrino spectrum. The various
cross-section ratios are given by
\begin{equation}
r=\frac{\int \int \int dE_{\nu }dTdT^{\prime }\phi _{B}(E_{\nu })\frac{%
d\sigma \left( \nu _{x}e\rightarrow \nu _{x}e\right) }{dT}(T,E_{\nu
})R(T,T^{\prime })}{\int \int \int dE_{\nu }dTdT^{\prime }\phi _{B}(E_{\nu })%
\frac{d\sigma \left( \nu _{e}e\rightarrow \nu _{e}e\right) }{dT}(T,E_{\nu
})R(T,T^{\prime })},
\end{equation}
\begin{equation}
\overline{r}_{e}=\frac{\int \int \int dE_{\nu }dTdT^{\prime }\phi
_{B}(E_{\nu })\frac{d\sigma \left( \overline{\nu }_{e}e\rightarrow \overline{%
\nu }_{e}e\right) }{dT}(T,E_{\nu })R(T,T^{\prime })}{\int \int \int dE_{\nu
}dTdT^{\prime }\phi _{B}(E_{\nu })\frac{d\sigma \left( \nu _{e}e\rightarrow
\nu _{e}e\right) }{dT}(T,E_{\nu })R(T,T^{\prime })},
\end{equation}
\begin{equation}
\overline{r}_{x}=\frac{\int \int \int dE_{\nu }dTdT^{\prime }\phi
_{B}(E_{\nu })\frac{d\sigma \left( \overline{\nu }_{x}e\rightarrow \overline{%
\nu }_{x}e\right) }{dT}(T,E_{\nu })R(T,T^{\prime })}{\int \int \int dE_{\nu
}dTdT^{\prime }(E_{\nu })\frac{d\sigma \left( \nu _{e}e\rightarrow \nu
_{e}e\right) }{dT}(T,E_{\nu })R(T,T^{\prime })},
\end{equation}
\begin{equation}
\overline{r}_{d}=\frac{\int \int \int dE_{\nu }dTdT^{\prime }\phi
_{B}(E_{\nu })\frac{d\sigma \left( \overline{\nu }_{x}d\rightarrow \overline{%
\nu }_{x}pn\right) }{dT}(T,E_{\nu })R(T,T^{\prime })}{\int \int \int dE_{\nu
}dTdT^{\prime }(E_{\nu })\frac{d\sigma \left( \nu _{e}d\rightarrow \nu
_{e}pn\right) }{dT}(T,E_{\nu })R(T,T^{\prime })}.
\end{equation}
From these relations, we obtain $r=0.15$, $\overline{r}_{x}=0.12$, $%
\overline{r}_{e}=0.19$ and $\overline{r}_{d}=0.95$.

It can be seen from Eqns. (8)-(10) that
\begin{equation}
\phi _{ES}=r\phi _{NC}+(1-r)\phi _{CC}+(\overline{r}_{e}-r\overline{r}%
_{d})\phi _{\overline{\nu }_{e}}-\left( r\overline{r}_{d}-\overline{r}%
_{x}\right) \phi _{\overline{\nu }_{x}}.
\end{equation}
In the absence of antineutrinos, we obtain
\begin{equation}
\phi _{ES}^{no~\overline{\nu}}=r\phi _{NC}+(1-r)\phi _{CC}
\end{equation}
which is the same as Eqn. (1). The effect of electron antineutrino component
in the solar boron neutrino flux is to increase the ES flux from $\phi
_{ES}^{no~\overline{\nu}}$while that of muon antineutrino component
is to decrease the ES flux from $\phi _{ES}^{no~\overline{\nu}}$.
Since, $\overline{\nu }_{e}$ flux is restricted to very small values by the
Super-Kamiokande \cite{9} and KamLAND \cite{15}, there will not be any
significant increase in the ES rate due to the contribution from the
electron antineutrinos. Hence, we neglect the $\overline{\nu }_{e}$ flux in
our analysis.

A value of ES flux smaller than $\phi _{ES}^{no~\overline{\nu}}$
(the ES flux in the absence of $\phi _{\overline{\nu }_{e}}$ and $\phi _{
\overline{\nu }_{x}}$) will be a signature for the muon antineutrino
component in the solar $^{8}$B neutrino flux. We calculate $\phi _{ES}^{no~\overline{\nu }}$ and its correlation coefficients with CC and NC
fluxes from Eqn. (16) using the CC and NC fluxes as given in Eqns. (2) and
(3) and the correlation coefficient between SNO CC and NC flux given by Eqn.
(5) and obtain
\begin{equation}
\phi _{ES}^{no~\overline{\nu}}=\left( 2.18\pm 0.09\right) \times
10^{6}cm^{-2}s^{-1},
\end{equation}
and
\begin{equation}
\rho \left( CC,ES\right) =0.91,
\end{equation}
\begin{equation}
\rho \left( ES,NC\right) =0.10.
\end{equation}
Thus, the SNO ES flux should be equal to $\left( 2.18\pm 0.09\right) \times
10^{6}$ cm$^{-2}$s$^{-1}$ and correlated with the CC and NC fluxes with
above correlation coefficients in case there are transitions into active
flavors only.

In case, we have independent measurements of CC and NC fluxes, they will be
correlated with the ES flux through the correlation coefficients
\begin{equation}
\rho \left( CC,ES\right) =\frac{(1-r)\sigma _{CC}}{\sqrt{r^{2}\sigma
_{NC}^{2}+(1-r)^{2}\sigma _{CC}^{2}}},
\end{equation}
\begin{equation}
\rho \left( ES,NC\right) =\frac{r\sigma _{NC}}{\sqrt{r^{2}\sigma
_{NC}^{2}+(1-r)^{2}\sigma _{CC}^{2}}},
\end{equation}
if there are transitions into active flavors only and no transitions into
antineutrinos. In the third $^{3}$He-phase of SNO, the CC and NC
measurements will be nearly independent. Then, the experimentally measured
ES fluxes should, not only, be related with the measured CC and NC fluxes
according to Eqn. (1), but also, it should be correlated with the CC and NC
fluxes with the correlation coefficients given by Eqns. (20) and (21).

The central value of the ES flux is clearly larger than the central value of
$\phi _{ES}^{no~\overline{\nu}}$. However, SNO ES flux has large
errors and includes $\phi _{ES}^{no~\overline{\nu}}$ within one
standard deviation. On the other hand, Super-Kamiokande ES flux has much
smaller errors. Since, there is very little difference between the various
cross-section ratios for Super-Kamiokande and SNO, we can use the
Super-Kamiokande ES flux in place of SNO ES flux and combine it with SNO CC
and NC fluxes. We treat the Super-Kamiokande ES flux as uncorrelated with
the SNO CC and ES fluxes.

The Super-Kamiokande ES rate \cite{4} is
\begin{equation}
\phi _{ES}^{SK}=\left( 2.35\pm 0.08\right) \times 10^{6} cm^{-2}s^{-1}
\end{equation}
and
\begin{equation}
\phi _{ES}^{SK}-\phi _{ES}^{no~\overline{\nu}}=\left( 0.17\pm
0.12\right) \times 10^{6} cm^{-2}s^{-1}
\end{equation}
which is non-zero at about 1.4 standard deviations. If there are transitions
into $\overline{\nu }_{x}$ in addition to the transitions into $\nu _{x}$
component, the ES flux should be smaller than $\phi _{ES}^{no~\overline{\nu}}$. But, we see that it is actually larger than $\phi _{ES}^{no~\overline{\nu}}$ upto 1.4$\sigma $. Thus, the Super-Kamiokande
data excludes any antineutrino production at 1.4$\sigma $ C.L.. However,
arbitrary admixtures of $\phi _{\nu _{x}}$ and $\ \phi _{\overline{\nu }
_{x}} $ are allowed above 1.4$\sigma $ C.L..

We eliminate $\phi _{\nu _{e}}$ from Eqns. (9) and (10) using Eqn. (8) to
obtain
\begin{equation}
\phi _{\nu _{x}}+\overline{r}_{d}\phi _{\overline{\nu }_{x}}=\phi _{NC}-\phi
_{CC}-\overline{r}_{d}\phi _{\overline{\nu }_{e}},
\end{equation}
\begin{equation}
r\phi _{\nu _{x}}+\overline{r}_{\mu }\phi _{\overline{\nu }_{x}}=\phi
_{ES}-\phi _{CC}-\overline{r}_{_{e}}\phi _{\overline{\nu }_{e}}.
\end{equation}
If the value of $\phi _{\overline{\nu }_{e}}$ is known, the above set of
linear equations can be solved for $\phi _{\nu _{x}}$and $\phi _{\overline{\nu }_{x}}$. The formal solutions of Eqns. (24) and (25) can be written as
\begin{equation}
\phi _{\overline{\nu }_{x}}=\frac{1}{r\overline{r}_{d}-\overline{r}_{x}}
\left[ r\phi _{NC}+(1-r)\phi _{CC}-\phi _{ES}+(\overline{r}_{e}-r\overline{r}
_{d})\phi _{\overline{\nu }_{e}}\right] ,
\end{equation}
\begin{equation}
\phi _{\nu _{x}}=\frac{1}{r\overline{r}_{d}-\overline{r}_{x}}\left[
\overline{r}_{d}\phi _{ES}-\overline{r}_{x}\phi _{NC}-(\overline{r}_{d}-
\overline{r}_{x})\phi _{CC}+\overline{r}_{d}(\overline{r}_{e}-\overline{r}
_{x})\phi _{\overline{\nu }_{e}}\right] .
\end{equation}
Since, $\phi _{\overline{\nu }_{e}}$ is restricted to very small values by
Super-Kamiokande \cite{9} and KamLAND \cite{15}, we neglect the $\phi _{
\overline{\nu }_{e}}$ term in the above equations. So,
\begin{equation}
\phi _{\overline{\nu }_{x}}=\frac{r\phi _{NC}+(1-r)\phi _{CC}-\phi _{ES}}{r
\overline{r}_{d}-\overline{r}_{x}},
\end{equation}
\begin{equation}
\phi _{\nu _{x}}=\frac{\overline{r}_{d}\phi _{ES}-\overline{r}_{x}\phi
_{NC}-(\overline{r}_{d}-\overline{r}_{x})\phi _{CC}}{r\overline{r}_{d}-
\overline{r}_{x}}.
\end{equation}
Since, $\phi _{ES}$ is larger than $\phi _{ES}^{no~\overline{\nu}}
=r\phi _{NC}+(1-r)\phi _{CC}$, the central value of muon antineutrino flux
will be negative and only upper bounds on it can be obtained. Substituting
the SNO fluxes from Eqns. (2)-(4) and using the correlation coefficients
from Eqns. (5)-(7), the active non-electronic antineutrino flux is found to
be
\begin{equation}
\phi _{\overline{\nu }_{x}}=(-7.0\pm 13.5)\times 10^{6} cm^{-2}s
^{-1}.
\end{equation}
The result is consistent with zero within one standard deviation.

If we use the Super-Kamiokande ES flux, given by Eqn. (22) instead of the
SNO ES flux we obtain
\begin{equation}
\phi _{\overline{\nu }_{x}}=(-7.6\pm 5.3)\times 10^{6}cm^{-2}s^{-1}.
\end{equation}
The upper bound on $\phi _{\overline{\nu }_{x}}$ is
\begin{equation}
\phi _{\overline{\nu }_{x}}<3.0\times 10^{6} cm^{-2}s^{-1}~~ at~ 2\sigma~C.L..
\end{equation}
It is found that the value of $\phi _{\overline{\nu }_{x}}$ becomes negative
below 1.4$\sigma $ C.L.. However, the data is not precise enough to restrict
the larger values of antineutrino flux. For instance, $\phi _{\overline{\nu }
_{x}}=5.69\times 10^{6}$cm$^{-2}$s$^{-1}$ is allowed at 2.5$\sigma $.

The total active neutrino flux is
\begin{equation}
\phi _{active}=\phi _{\nu _{e}}+\phi _{\nu _{x}}+\phi _{\overline{\nu
}_{x}}.
\end{equation}
Substituting the values of $\phi _{\nu _{e}}$, $\phi _{\nu _{x}}$ and $\phi
_{\overline{\nu }_{x}}$ from Eqns. (8), (29) and (28) in Eqn. (33), we obtain

\begin{equation}
\phi _{active}=\frac{1}{r\overline{r}_{d}-\overline{r}_{x}}\left[
\left( r-\overline{r}_{x}\right) \phi _{NC}+\left( 1-\overline{r}_{d}\right)
\left( \left( 1-r\right) \phi _{CC}-\phi _{ES}\right) \right] .
\end{equation}
We note that in the limit $\overline{r}_{d}\sim 1$, $\phi _{ES}\sim r\phi
_{NC}+(1-r)\phi _{CC}$ and so $\phi _{active}\sim \phi _{NC}$. The
sterile flux can, thus, be calculated from the relation
\begin{equation}
\phi _{sterile}=\phi _{B}^{SSM}-\phi _{active},
\end{equation}
where the SSM prediction for the total boron neutrino flux is \cite{16}
\begin{equation}
\phi _{B}^{SSM}=\left( 5.69\pm 0.91\right) \times 10^{6} cm^{-2}s^{-1}.
\end{equation}
Using the SNO CC and NC fluxes alongwith the Super-Kamiokande ES flux,
sterile neutrino flux is found to be
\begin{equation}
\phi _{sterile}=1.25\pm 0.99\times 10^{6} cm^{-2}s^{-1},
\end{equation}
which is non-zero at about $1.26$ standard deviations and gives a 2$\sigma $
C.L$.$ upper bound of
\begin{equation}
\phi _{sterile}<3.23\times 10^{6} cm^{-2}s^{-1}.
\end{equation}

In order to compare our results with those obtained by Chauhan and Pulido
\cite{12}, we calculate the ratio of non-electronic neutrino flux to the
total active (neutrino+antineutrino) flux:
\begin{equation}
\sin ^{2}\psi =\frac{\phi _{\nu _{x}}}{\phi _{\nu _{x}}+\phi _{\overline{\nu
}_{x}}}.
\end{equation}
On substituting $\phi _{\nu _{x}}$ and $\phi _{\overline{\nu }_{x}}$ from
Eqns. (28) and (29) in the above equation, we obtain
\begin{equation}
\sin ^{2}\psi =\frac{\overline{r}_{d}\gamma -\overline{r}_{x}}{r-\overline{r}%
_{x}-\left( 1-\overline{r}_{d}\right) \gamma }
\end{equation}
where
\begin{equation}
\gamma =\frac{\phi _{ES}-\phi _{CC}}{\phi _{NC}-\phi _{CC}}.
\end{equation}
The analysis of Ref. [12] is based upon Eqn. (40). Substituting the SNO CC
and NC fluxes and Super-Kamiokande ES flux in Eqn. (41) and taking into
account the anticorrelation between the SNO CC and NC fluxes, we obtain
\begin{equation}
\gamma =0.20\pm 0.04.
\end{equation}
Using this value of $\gamma $, we can calculate the value of sin$^{2}\psi $
from Eqn. (40) to obtain
\begin{equation}
\sin ^{2}\psi =3.5\pm 2.2
\end{equation}
which is 1.\thinspace 1$\sigma $ above unity. Hence, no
neutrino-antineutrino admixture is allowed upto 1.\thinspace 1$\sigma $.
However, arbitrary neutrino-antineutrino admixtures are allowed in the range
1.1- 1.6$\sigma $. For example,

\begin{equation}
\sin ^{2}\psi =0.5~~ at~ 1.36 \sigma~ C.L.
\end{equation}
Our results differ from those obtained by Chauhan and Pulido \cite{12}
because we incorporate the Super-Kamiokande flux in the analysis from the
very start instead of putting it as a constraint at the end and, also,
because we have taken into account the anticorrelation between the SNO CC
and NC fluxes.

Now, we consider a special case in which there are transitions only into
active antineutrinos and no transitions into active neutrinos. This will
allow us to obtain the maximum possible antineutrino flux allowed by the
data. In this scenario, we have

\begin{equation}
\phi _{\overline{\nu }_{x}}=\frac{\phi _{NC}-\phi _{CC}}{\overline{r}_{d}}
\end{equation}
which gives
\begin{equation}
\phi _{\overline{\nu }_{x}}=\left( 3.26\pm 0.41\right) \times 10^{6} cm^{-2}s^{-1}.
\end{equation}
Thus, the 2$\sigma $ upper bound on the antineutrino flux is
\begin{equation}
\phi _{\overline{\nu }_{x}}<4.08\times 10^{6} cm^{-2}s^{-1}.
\end{equation}

In conclusion, a non-electronic antineutrino component in the solar boron
neutrino flux is ruled out by the latest 391-Day Salt Phase SNO Data Set and
the Super-Kamiokande ES measurement at 1.4$\sigma $ C.L.. However, arbitrary
$\nu _{x}-\overline{\nu }_{x}$ admixtures are still allowed above this
confidence level. The upper bound on muon/tauon antineutrino flux coming
from the Sun is $\phi _{\overline{\nu }_{x}}<3.0\times 10^{6}$cm$^{-2}$s$^{-1}$ at 2$\sigma $ C.L.. With the future precision measurements of the CC
and NC fluxes at SNO and ES flux at Super-Kamiokande, it will be possible to
further constrain the $\nu _{x}-\overline{\nu }_{x}$ admixtures present in
the solar boron neutrino flux to smaller values. Combining the
Super-Kamiokande ES flux with the SNO CC and NC fluxes, the sterile flux is
found to be non-zero at 1.26 standard deviations. The upper bound on the
sterile flux is 3.23$\times $10$^{6}$cm$^{-2}$s$^{-1}$ at 2$\sigma $ C.L.

\bigskip

{\LARGE Acknowledgments}

\medskip

One of the authors (S. K.) acknowledges the financial support provided by
Council for Scientific and Industrial Research (CSIR), Government of India.
The research work of the other two authors (S. D. and S. V.) is supported by
the Board of Research in Nuclear Sciences (BRNS), Department of Atomic
Energy, Government of India {\it vide} Grant No. 2004/37/23/BRNS/399.

\end{document}